\documentclass[aps,twocolumn,superscriptaddress,showpacs]{revtex4}

\usepackage[dvips]{graphicx} 
\usepackage{color}           
\usepackage{amsmath,amssymb,amsfonts}
\usepackage{bm}
\newcommand{\nc}{\newcommand}           
\nc{\vc}[1]     {\mbox{\boldmath $#1$}} 
\nc{\mapleft}[1]{                       
 \smash{\mathop{                      %
  \hbox to 0.90cm{\rightarrowfill} }\limits_{#1}}}

\nc{\beq}     {\begin{eqnarray}}
\nc{\eeq}    {\end{eqnarray}}
\nc{\bra}       {\langle}               
\nc{\ket}       {\rangle}               
\nc{\bras}[1]   {\langle#1|}            
\nc{\kets}[1]   {|#1\rangle}            
\nc{\del}       {\partial}              

\newcommand{\lw}[1]{\smash{\lower1.75ex\hbox{#1}}}

\nc{\red}[1]    {\textcolor{red}{#1}}  

\nc{\mydraft}	{\setlength{\topmargin}{-1.5cm}}
\mydraft
\begin{document}

\title{New many-body method using cluster expansion diagrams \\ with tensor-optimized antisymmetrized molecular dynamics}

\author{Takayuki Myo\footnote{takayuki.myo@oit.ac.jp}}
\affiliation{General Education, Faculty of Engineering, Osaka Institute of Technology, Osaka, Osaka 535-8585, Japan}
\affiliation{Research Center for Nuclear Physics (RCNP), Osaka University, Ibaraki, Osaka 567-0047, Japan}

\author{Mengjiao Lyu}
\affiliation{College of Science, Nanjing University of Aeronautics and Astronautics, Nanjing 210016, China}

\author{Hiroshi Toki}
\affiliation{Research Center for Nuclear Physics (RCNP), Osaka University, Ibaraki, Osaka 567-0047, Japan}

\author{Hisashi Horiuchi}
\affiliation{Research Center for Nuclear Physics (RCNP), Osaka University, Ibaraki, Osaka 567-0047, Japan}

\author{Qing Zhao}
\affiliation{School of Science, Huzhou University, Huzhou 313000, Zhejiang, China}

\author{Masahiro Isaka}
\affiliation{Hosei University, Chiyoda-ku, Tokyo 102-8160, Japan}

\author{Hiroki Takemoto}
\affiliation{Faculty of Pharmacy, Osaka Medical and Pharmaceutical University, Takatsuki, Osaka 569-1094, Japan} 

\author{Niu Wan}
\affiliation{School of Physics and Optoelectronics, South China University of Technology, Guangzhou 510641, China}

\date{\today}

\begin{abstract}%
  We propose a new many-body method based on the correlation functions, in which
  the multiple products of the correlation functions are expanded into the many-body diagrams using the cluster expansion method
  and every diagram is independently optimized in the total-energy variation.
  We apply this idea to the tensor-optimized antisymmetrized molecular dynamics (TOAMD) using the bare nucleon-nucleon interaction 
  and show the results of the $s$-shell nuclei within the triple products of the correlation functions of tensor and central-types. 
  We evaluate the effect of the independent optimization of the many-body diagrams on the solutions.
  It is found that the triple products provides the sizable effect in the present scheme,
  which results in the good reproduction of the total energy and the Hamiltonian components of nuclei
  with respect to the few-body calculations.
\end{abstract}

\pacs{
21.60.Gx, 
21.30.-x  
}
\maketitle

\section{Introduction}
It is known that the bare nucleon--nucleon ($NN$) interaction has the strong characteristics of the short-range repulsion and the tensor force \cite{pieper01,aoki13}.
In finite nuclei, the short-range repulsion reduces the amplitudes of nucleon pairs at short distance as the short-range correlation,
and the tensor force produces the $D$-wave transition of nucleon pairs as the tensor correlation.
These correlations are physically different, but commonly induce the high-momentum motion of nucleons in nuclei \cite{schiavilla07}. 

In light-mass region, the Green's function Monte Carlo (GFMC) method has been developed to treat the nucleon-nucleon correlations directly in nuclei.
In GFMC, they can reproduce the binding energies and the energy spectra of light nuclei up to $^{12}$C within three-nucleon forces \cite{carlson15}.
It is also shown that the one-pion exchange force produces about 80\% of the entire attraction of the two-body interaction energy~\cite{pieper01}.
The pion exchange process is a dominant source of the tensor force, hence the treatment of tensor force is essential to describe the nuclear many-body systems.
In GFMC, the correlation functions are multiplied to the reference nuclear wave function
and it is important how to optimize the correlation functions appropriately to minimize the total energy of nuclei.

Based on the approach using the correlation functions, we have developed a new many-body method for finite nuclei treating strong interaction directly,
so-called the tensor-optimized antisymmetrized molecular dynamics (TOAMD) \cite{myo15,myo17a,myo17b,myo17c,myo17d,myo17e,myo18,lyu18a,lyu18b,zhao19,lyu20,myo21}.
The TOAMD framework is a variational method of describing the correlations induced by nuclear force.
In TOAMD, we use the AMD wave function as a reference state \cite{kanada03}, which can describe the structures of light nuclei with effective interactions.
In TOAMD, we adopt the two-body correlation functions of tensor-- and central--operator types corresponding to the $NN$ interaction in nuclei, and they are multiplied to the AMD reference state.
We further consider the multiple products of these correlation functions, which finally define the total nuclear wave function involving the many-body correlations.
We can determine each of the correlation functions in these multiple products using the total-energy variation.

So far, we have applied TOAMD to the $s$-shell nuclei and the $p$-shell $^5$He nucleus within the double products of the correlation functions \cite{myo17a,myo17b,myo17c,myo17d,myo21}, which is a second order of the correlation functions. 
In TOAMD, the multiple products of the correlation functions are expanded into the series of the many-body diagrams in the cluster expansion.
In the expansion of the specific products of the correlation functions, 
the relative weights between many-body diagrams are fixed by the particle exchange properties of the diagrams, which is so-called the symmetry factors.
We take all diagrams fixing the relative weights, while forms of the correlation functions are optimized in the individual products to minimize the total energy of a nucleus, which increases the variational accuracy of TOAMD.
Owing to this property, we have shown that the binding energies in TOAMD are better than those in the Jastrow method using common form of the correlation functions for every pair \cite{myo17b,myo17c,jastrow55}.
For $^5$He, we further performed the generator coordinate calculation by superposing several AMD reference states,
and discussed the excited state with the possibility of the $^3$He+$d$ clustering structure,
which is different from the ground state with a $^4$He+$n$ type.
We recently adopt the concept of TOAMD in the nuclear matter calculation with the bare $NN$ interactions \cite{myo19,yamada19,wan20}.

In TOAMD, we can extend the variational accuracy by adding the higher orders of the correlation functions in the total wave function,
in which the matrix elements at any order of TOAMD are presented in the analytical form.
In the present study, we extend TOAMD in two aspects:
1) We increase the order of TOAMD from the second to the third one, which includes the triple products of the correlation functions.
2) We use the cluster expansion to decompose the multiple products of the correlation functions into the many-body diagrams.
We extend this process by treating each diagram as the independent basis state, and the correlation functions in each diagram are determined variationally. This extension further increases the variational space of TOAMD. 
In this study, we confirm the effects of these new extensions in the calculation of $s$-shell nuclei with the bare $NN$ interactions.

In Sec.~\ref{sec:method}, we explain the many-body framework of TOAMD and propose a new treatment of the cluster-expansion diagrams. 
In Sec.~\ref{sec:results}, we show the results of $s$-shell nuclei $^3$H and $^4$He with the bare Argonne-type $NN$ interactions
and discuss the effect of the present new method.
A summary is given in Sec.~\ref{sec:summary}.

\section{Method}\label{sec:method}

\subsection{Tensor-optimized antisymmetrized molecular dynamics (TOAMD)}\label{sec:TOAMD}
We explain the wave function of TOAMD for the nucleus with mass number $A$.
We start from the AMD wave function $\Phi_{\rm AMD}$, which is the Slater determinant of the nucleon wave functions $\phi_{\sigma\tau}(\vc r)$, as
\begin{eqnarray}
\Phi_{\rm AMD}
&=& \frac{1}{\sqrt{A!}}\, {\rm det} \left\{ \prod_{i=1}^A \phi_{\sigma_i\tau_i}(\vc{r}_i) \right\}~,
\label{eq:AMD}
\\
\phi_{\sigma\tau}(\vc{r})&=&\left(\frac{2\nu}{\pi}\right)^{3/4} e^{-\nu(\bm{r}-\bm{D})^2} \chi_{\sigma} \chi_{\tau}.
\label{eq:Gauss}
\end{eqnarray}
The function $\phi_{\sigma\tau}(\vc r)$ consists of a Gaussian wave packet with a range parameter $\nu$ and a centroid position $\vc D$, 
a spin part $\chi_{\sigma}$ and an isospin part $\chi_{\tau}$.
In this work, $\chi_{\sigma}$ is the up or down component and $\chi_{\tau}$ is proton or neutron.

In TOAMD, we adopt two kinds of the two-body correlation functions to make the correlated wave function from $\Phi_{\rm AMD}$;
one is $F_D$ for the tensor force with a relative $D$-wave transition and the other is $F_S$ for short-range repulsion.
These functions are defined explicitly as
\begin{eqnarray}
F_D
&=& \sum_{t=0}^1\sum_{i<j}^A f^{t}_{D}(r_{ij})\, S_{12}(\bm{\hat{r}}_{ij}) \left(P_{\tau,ij}\right)^t \,,
\label{eq:Fd}
\\
F_S
&=& \sum_{t=0}^1\sum_{s=0}^1 \sum_{i<j}^A f^{t,s}_{S}(r_{ij}) \left( P_{\tau,ij}\right)^t \left(P_{\sigma,ij}\right)^s \,,
\label{eq:Fs}
\end{eqnarray}
with a relative coordinate $\vc r_{ij}=\vc r_i - \vc r_j$.
The operators $P_{\tau}$ and $P_{\sigma}$ exchange the isospin and spin components between nucleons, respectively.
The labels $t$ and $s$ represent the spin-isospin dependence of the correlation functions.
The pair functions $f^{t}_{D}(r)$ and $f^{t,s}_{S}(r)$ are determined in the total-energy variation.
The functions $F_D$ and $F_S$ express the correlations between two nucleons in nuclei.
It is noted that $F_S$ generally describes the central correlation including short-range one.

We further consider the multiple products of the correlation functions to make the many-body correlations beyond two-body one,
and these terms are multiplied to $\Phi_{\rm AMD}$ and superposed.
We define the TOAMD wave function starting from the single correlation functions as the first order:
\begin{eqnarray}
\Phi_{\rm TOAMD}^{\rm single}
&=& (1+F_S+F_D) \times\Phi_{\rm AMD}\,.
\label{eq:TOAMD1}
\end{eqnarray}
We introduce the next second order of the correlation functions in TOAMD by successively adding the double products consisting of $F_D$ and $F_S$ as
\begin{eqnarray}
\Phi_{\rm TOAMD}^{\rm double}
&=& (1+F_S+F_D+F_S F_S+F_D F_S+F_D F_D)
\nonumber\\
&\times&\Phi_{\rm AMD}~.
\label{eq:TOAMD2}
\end{eqnarray}
We finally define the third order including the triple products of the correlation functions as
\begin{eqnarray}
\Phi_{\rm TOAMD}^{\rm triple}
&=& (1+F_S+F_D+F_S F_S+F_D F_S+F_D F_D
\nonumber\\
&+& F_D F_S F_S + F_D F_D F_S)\times \Phi_{\rm AMD}~.
\label{eq:TOAMD3}
\end{eqnarray}
Here we add the terms having $F_D$ and $F_S$ simultaneously in Eq.~(\ref{eq:TOAMD3}), because
it is found that both correlations are important in the second order analysis \cite{myo17a,myo17d}.
For other combinations of triple products, $F_S F_S F_S$ and $F_D F_D F_D$, each term has the overlap
with the $F_D F_D F_S$ and $F_D F_S F_S$, respectively, because two $D$-waves of $F_D F_D$ can be coupled with two $S$-waves of $F_S F_S$,
hence the partial effects of $F_S F_S F_S$ and $F_D F_D F_D$ are included in $F_D F_D F_S$ and $F_D F_S F_S$, respectively.
Numerically, the $F_D F_D F_D$ term requires the largest computational effort because of the triple products of the tensor operators.
From these reasons, in the present analysis, which is a first one with the triple correlation functions in TOAMD,
we put a priority on the terms including both of $F_D$ and $F_S$. 

We successively add the terms of multiple products of the correlation functions starting from $\Phi_{\rm AMD}$
to confirm the convergence of the solutions.
These three kinds of the TOAMD wave function corresponds to the power series expansion using $F_D$ and $F_S$,
but all the functions of $F_D$ and $F_S$ in each term of Eqs.~(\ref{eq:TOAMD1}), (\ref{eq:TOAMD2}) and (\ref{eq:TOAMD3}) can be different and their functional forms are variationally determined,
although we use the common notations of $F_D$ and $F_S$.
This property of TOAMD increases the variational degrees of freedom in comparison with the Jastrow method,
in which the common correlation functions are assumed for all nucleon-pairs \cite{jastrow55}.
In the present study, we adopt up to the third order of the TOAMD wave function in Eq.~(\ref{eq:TOAMD3}).

\begin{figure}[b]
\centering
\includegraphics[width=8.5cm,clip]{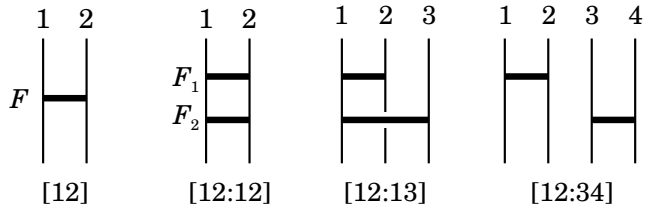}
\caption{Diagrams of the cluster expansion of $F^2$.
  The vertical lines are the particles and the horizontal lines are the correlation functions $F$.
  The numbers in the square brackets such as [12:12] represent the configurations to specify the diagram \cite{myo17c},
  in which each $F$ is separated by column.}
\label{fig:FF2}
\end{figure}

We explain the Hamiltonian with a two-body realistic $NN$ interaction as
\begin{eqnarray}
    H
&=& \sum_i^{A} t_i - T_{\rm G} + \sum_{i<j}^{A} v_{ij}\, .
    \label{eq:Ham}
\end{eqnarray}
Here, $t_i$ and $T_{\rm G}$ are the kinetic energies of each nucleon and the center-of-mass, respectively.
We employ the realistic $NN$ interaction $v_{ij}$ of the AV6$^\prime$ and AV8$^\prime$ potentials \cite{wiringa95,pudliner97,wiringa02} consisting of central and tensor terms and the $LS$ term is added in the AV8$^\prime$ potential.
We also add the point Coulomb interaction for protons.
The total energy $E$ of a nucleus is given as
\begin{eqnarray}
    E
&=&\frac{\langle\Phi_{\rm TOAMD} |H|\Phi_{\rm TOAMD}\rangle}{\langle\Phi_{\rm TOAMD} |\Phi_{\rm TOAMD}\rangle}
    \nonumber
    \\
&=&\frac{\langle\Phi_{\rm AMD} | \widetilde{H} |\Phi_{\rm AMD}\rangle}{\langle\Phi_{\rm AMD} | \widetilde{N} |\Phi_{\rm AMD}\rangle}.
\label{eq:E_TOAMD}
\end{eqnarray}
We expand the TOAMD wave function $\Phi_{\rm TOAMD}$ using Eq.~(\ref{eq:TOAMD3}) and
define the correlated operators with the upper tilde in Eq.~(\ref{eq:E_TOAMD}) with the AMD wave function $\Phi_{\rm AMD}$. 
The operators $\widetilde{H}$ and $\widetilde{N}$ are the summation of the multiple products of the operators such as $F^\dagger H F$ and $F^\dagger F$, with $F$ being $F_D$ or $F_S$. 
Each of the multiple products of the correlation functions in $\widetilde{H}$ and $\widetilde{N}$
is expanded into the irreducible many-body operators using the cluster expansion \cite{myo15,myo17d},
in which each of many-body operator is expressed by using the specific configuration and diagram.
In the simplest case of $F^\dagger F$, this term is expanded into two-body, three-body and four-body operators,
which are expressed with the configurations of [12:12], [12:13] and [12:34] with particle indices in the square brackets, respectively.
The corresponding diagrams are displayed in Fig. \ref{fig:FF2}.
For two-body interaction $V$, the correlated interaction $F^\dagger V F$ provides diagrams
from the two-body term as [12:12:12] to the six-body term as [12:34:56].
For one-body operators such as the kinetic energy, the correlated operators are written by diagrams with up to the five-body term as [12:3:45].

In TOAMD, we adopt all the terms in the cluster expansion of the correlated operators in Eq.~(\ref{eq:E_TOAMD}), 
and calculate the corresponding matrix elements with the AMD wave function.
Due to this condition, TOAMD becomes a variational method for nuclei with the bare $NN$ interaction.
The TOAMD wave function has two kinds of variational functions; the AMD wave function $\Phi_{\rm AMD}$ and the correlation functions $F_D$ and $F_S$.
We use the variational principle with respect to the total energy $E$ as $\delta E=0$ in Eq.~(\ref{eq:E_TOAMD}).
In the determination of the radial forms of $F_D$ and $F_S$,
we express the pair functions $f^{t}_{D}(r)$ and $f^{t,s}_{S}(r)$ in the linear combination of the Gaussian functions
with the number $N_{\rm G}$ as
\begin{eqnarray}
   f^t_D(r)
&=&\sum_{n=1}^{N_{\rm G}} C^t_n \, e^{-a^t_n r^2},
   \label{eq:cr1}
   \\
   f^{t,s}_S(r)
&=&\sum_{n=1}^{N_{\rm G}} C^{t,s}_n\, e^{-a^{t,s}_n r^2}.
   \label{eq:cr2}
\end{eqnarray}
The variational parameters are $a^t_n$, $a^{t,s}_n$, $C^t_n$ and $C^{t,s}_n$ with the index $n$. 
We set $N_{\rm G}=7$ at most to get the converging solutions.
For the Gaussian ranges $a^t_n$, $a^{t,s}_n$, we search for the optimal values in a wide range.
The expansion coefficients $C^t_n$ and $C^{t,s}_n$ are obtained in the eigenvalue problem of the Hamiltonian matrix.
For the double and triple products of the correlation functions such as $F_D F_D$,
the products of Gaussian functions in Eqs. (\ref{eq:cr1}) and (\ref{eq:cr2}) are treated as the single basis functions.

Finally, we define the TOAMD wave function in a linear combination form using the coefficients $\widetilde{C}_\alpha$ for the Gaussian expansion of the correlation functions
\begin{eqnarray}
   \Phi_{\rm TOAMD}
&=& \sum_{\alpha=0} \widetilde{C}_\alpha\,  \Phi_{{\rm TOAMD},\alpha} \,,
   \label{eq:linear}
\end{eqnarray}
where the label $\alpha$ is the set of the Gaussian index $n$ and the labels $s$ and $t$ in the correlation functions.
The summation includes the basis states with single, double and triple correlations.
We give the AMD wave function the label of $\alpha=0$.
The Hamiltonian and norm matrices are $H_{\alpha \beta}$ and $N_{\alpha \beta}$, respectively, and are given as
\begin{eqnarray}
   H_{\alpha \beta}
&=& \langle\Phi_{{\rm TOAMD},\alpha} |H|\Phi_{{\rm TOAMD},\beta}\rangle
   \nonumber\\
   \label{eq:Ham_crr}
&=& \langle\Phi_{\rm AMD} |\widetilde{H}_{\alpha \beta}|\Phi_{\rm AMD}\rangle\, ,
\\
   N_{\alpha \beta}
&=& \langle\Phi_{{\rm TOAMD},\alpha} |\Phi_{{\rm TOAMD},\beta}\rangle
   \nonumber\\
&=& \langle\Phi_{\rm AMD}|\widetilde{N}_{\alpha \beta} |\Phi_{\rm AMD}\rangle\, .
   \label{eq:Nrm_crr}
\end{eqnarray}
We solve the following generalized eigenvalue problem, and obtain the total energy $E$ and the coefficients $\widetilde{C}_\alpha$.
\begin{eqnarray}
   \sum_{\beta=0} \left( H_{\alpha\beta} - E\, N_{\alpha\beta} \right) \widetilde{C}_\beta &=&0.
   \label{eq:eigen}
\end{eqnarray}

  It is noted that the expression in Eq.(\ref{eq:linear}) is general for all the terms of the basis states in TOAMD including the double and triple products
  of the correlation terms.
  By solving Eq.~(\ref{eq:eigen}), we optimize all the correlation functions simultaneously with the double and triple products.
  This means that when we successively add the correlation terms, all the terms are optimized at each step of the calculation under the variational principle.

We explain the procedure to evaluate the matrix elements of the correlated many-body operators using the AMD wave function in Eqs.~(\ref{eq:Ham_crr}) and (\ref{eq:Nrm_crr}).
We express the $NN$ interaction $v_{ij}$ as a sum of Gaussians, similarly to the correlation functions $f_{ij}$ and then we essentially perform the Gaussian integration.
In the cluster expansion of $\widetilde{H}_{\alpha \beta}$ and $\widetilde{N}_{\alpha \beta}$,
the many-body operators have the set of the square of the interparticle coordinates $\vc r_{ij}^{\,2}$ in the Gaussians
with various connections such as those shown in Fig.~\ref{fig:FF2}.
We perform the Fourier transformation of each Gaussian in $v_{ij}$ and $f_{ij}$ with the momentum $\vc k$,
which becomes the products of the plane waves, $e^{i\bm{k}\cdot \bm{r}_i}\, e^{-i\bm{k}\cdot \bm{r}_j}$,
and we calculate the single-particle matrix elements of the plane waves in AMD using Eq.~(\ref{eq:Gauss}).
We finally perform the multiple integration over all momenta and obtain the matrix elements of the many-body operators
in the analytical form \cite{myo15,myo17d}.

\subsection{New cluster-expansion}\label{sec:OPT}
We explain a new scheme of the cluster expansion of the multiple products of the correlation functions in TOAMD.
In the TOAMD basis states, we have the terms of $F^2$ and $F^3$, which produce the irreducible many-body diagrams in the cluster expansion. 
We treat all the resulting diagrams independently and determine the correlation functions in each diagram variationally.
This is a new idea of the present study and we call this new framework ''New-TOAMD''.
We explain New-TOAMD in more detail and for simplicity,
we omit the spin-isospin dependence in the two-body correlation function $F$.
We start from the expression of the correlation function $F$ as
\begin{eqnarray}
   F&=& \sum^A_{i<j} f_{ij}~=~\frac12 \sum^A_{i\neq j} f_{ij}~\Leftrightarrow~\frac12 [12],
   \label{eq:F}
   \\
   f_{ij} &=& \sum_n^{N_G} c_n\,g^n_{ij},\qquad g^n_{ij}~=~e^{-a_n(\bm{r}_i-\bm{r}_j)^2},
   \\
   F&=& \sum_n^{N_G} c_n\,G_n, \qquad G_n~=~\frac12 \sum^A_{i,j} g^n_{ij}.
\end{eqnarray}
The factor $\dfrac12$ in Eq.~(\ref{eq:F}) is a symmetry factor to control the pair number in the correlation consistently.
The left-right arrow transforms the correlation into the configuration with square brackets.
The function $g^n$ is a Gaussian with a range index $n$ for one nucleon-pair and $G_n$ is the summation of $g^n$ over all pairs.
In the summation, we simply denote the condition of $\sum_{i\neq j}^A$ as $\sum_{i,j}^A$.
In the same manner, for the $F^2$ case, 
\begin{eqnarray}
  FF' &=& \left( \frac12 \sum^A_{i,j} f_{ij} \right) \cdot \left( \frac12 \sum^A_{i,j} f'_{ij} \right)
  \\
  &=& \frac12 \sum^A_{i,j} f_{ij}f'_{ij} + \sum^A_{i,j,k} f_{ij} f'_{ik} + \frac14 \sum^A_{i,j,k,l} f_{ij} f'_{kl}
\\
&=& \sum_{n}^{N_G} \sum_{n'}^{N_G} c_n c_{n'}
\label{eq:F2}\\
&\times&
\left(
\frac12 \sum^A_{i,j} g^n_{ij}g^{n'}_{ij} + \sum^A_{i,j,k} g^n_{ij} g^{n'}_{ik} + \frac14 \sum^A_{i,j,k,l} g^n_{ij} g^{n'}_{kl}
\right),
\nonumber\\
  &\Leftrightarrow& \frac12 [12:12] + [12:13] + \frac14 [12:34].
\end{eqnarray}
The diagrams are shown in Fig. \ref{fig:FF2}.
In the ordinary TOAMD, we keep the relation of weights, namely, symmetry factors, among two-, three- and four-body diagrams,
which indicates that the form of $F^2$ is also kept in the wave function.
The basis number of the $F^2$ term is $N_G^2$ and the products of $c_n c_{n'}$ are variational coefficients.

It is noticed that each of many-body diagrams has the symmetry with respect to the particle exchange.
This means that each diagram can be the single basis state and 
it is not necessary to fix the relative weights of three diagrams using the symmetry factors in Eq. (\ref{eq:F2}).
In New-TOAMD, the weight of each diagram can be determined in the total-energy variation as
\begin{eqnarray}
  FF' 
&\Rightarrow& \sum_{n}^{N_G} \sum_{n'}^{N_G} 
\left( c_{n,n'}^{[2]} G^{[2]}_{n,n'} + c_{n,n'}^{[3]} G^{[3]}_{n,n'} + c_{n,n'}^{[4]} G^{[4]}_{n,n'}
\right),
\nonumber
\\
G^{[2]}_{n,n'} &=& \sum^A_{i,j} g^n_{ij}g^{n'}_{ij}\qquad \mbox{for 2-body correlation},
\\
G^{[3]}_{n,n'} &=& \sum^A_{i,j,k} g^n_{ij} g^{n'}_{ik}\qquad \mbox{for 3-body correlation},
\\
G^{[4]}_{n,n'} &=& \sum^A_{i,j,k,l} g^n_{ij} g^{n'}_{kl}\qquad \mbox{for 4-body correlation}.
\label{eq:F3}
\end{eqnarray}
We treat each diagram independently in the total wave function with the corresponding weights $c_{n,n'}^{[2]}$,
$c_{n,n'}^{[3]}$, and $c_{n,n'}^{[4]}$, for two-body, three-body, and four-body diagrams, respectively. 
Hence the basis number of the $F^2$ term increases as $3\times N_G^2$ and we can evaluate the physical contributions of each diagram in nuclei.
It is noted that in this new scheme the form of $F^2$ is not necessary to be kept in the wave function,
because we do not keep the symmetry factors of the expanded diagrams.
For the $F^3$ terms, we perform the similar extension of the cluster expansion diagrams, explicitly shown later.

\begin{figure}[th]
\centering
\includegraphics[width=7.0cm,clip]{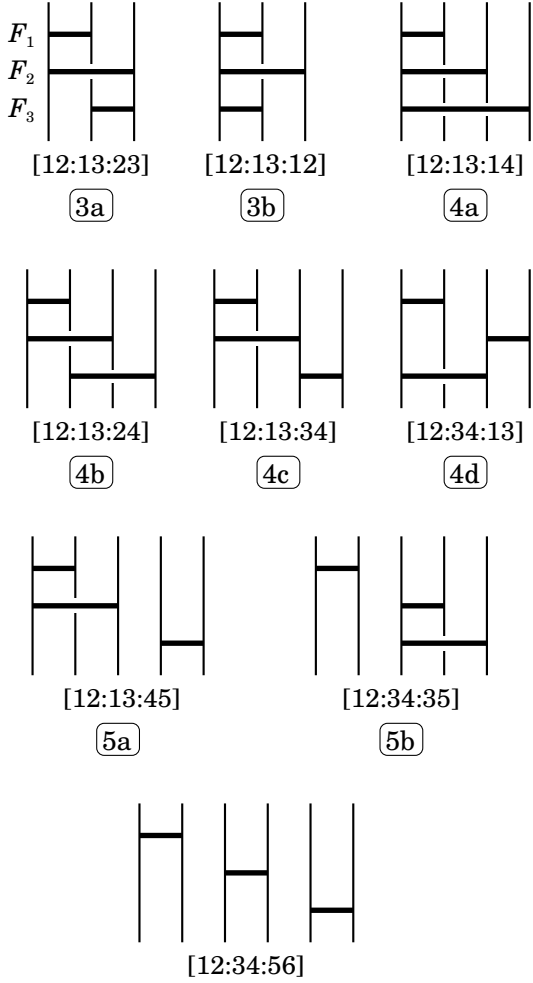}
\caption{Diagrams of the cluster expansion of $F^3$ included in the present calculation.}
\label{fig:FFF2}
\end{figure}

\begin{figure}[th]
\centering
\includegraphics[width=7.6cm,clip]{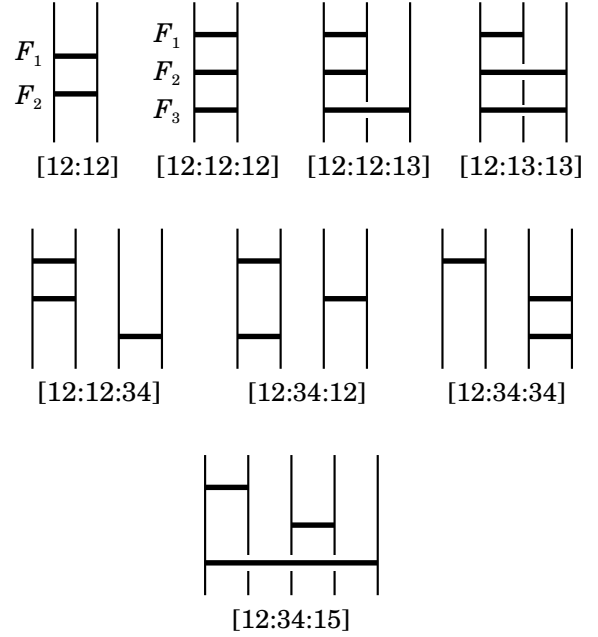}
\caption{The omitted diagrams of the cluster expansion of $F^2$ and $F^3$ because of the duplicated condition.}
\label{fig:FFF3}
\end{figure}

From the property of the independent treatment of the many-body diagrams,
we can also omit the specific diagrams, the effect of which can be expressed by other diagrams. 
For example, [12], [12:12], and [12:12:12] are the ladder diagrams of the cluster expansion of $F$, $F^2$ and $F^3$, respectively,
as shown in Fig. \ref{fig:FF2}.
In the explicit forms, they are given as
\begin{eqnarray}
\sum_{i,j} f_{ij} &\Leftrightarrow&  \dfrac12 [12],
\\
\sum_{i,j} f_{ij} f'_{ij} &\Leftrightarrow& \dfrac12  \left[12:12\right],
\\
\sum_{i,j} f_{ij} f'_{ij} f''_{ij} &\Leftrightarrow& \dfrac12  \left[12:12:12\right] .
\end{eqnarray}
They can give the same effect of two-body correlations in the wave function.
Hence it is sufficient to include the diagram [12] at least in the total wave function,
and we can remove the diagrams of [12:12] and [12:12:12].
We adopt this new scheme of the cluster expansion in New-TOAMD.

In the present study, we consider up to the $F^3$ terms in TOAMD and
some of the diagrams are omitted because of the duplicated correlations already included in the wave function.
In the expansion of $F^2$, we adopt [12:13] and [12:34] as shown in Fig. \ref{fig:FF2}. 
In the $F^3$ case, for example, the three-body diagram of [12:12:13] has partly the ladder diagram of [12:12] and then 
can be omitted because of the diagram of [12:13] in the $F^2$ term. 
In Fig. \ref{fig:FFF2}, we summarize the many-body diagrams in the cluster expansion of $F^3$,
which are included in the calculation with the independent weights.
We put the labels of 3a and 3b for three-body diagrams, 4a,\,4b,\,4c and 4d for four-body case, and 5a,\,5b for five-body case,
which are used to identify each diagram in the analysis.
In Fig. \ref{fig:FFF3}, we also show the diagrams, which are omitted in the calculation
because of the duplicated condition associated with other diagrams shown in Figs.\ref{fig:FF2} and \ref{fig:FFF2}.

  We explain the variational parameters in New-TOAMD.
  In the AMD wave function $\Phi_{\rm AMD}$, the range parameter $\nu$ and the centroid position $\vc D$ in Eq.~(\ref{eq:Gauss})
  are determined to minimize the total energy.
  For correlation functions with Gaussian expansion in Eqs. (\ref{eq:cr1}) and (\ref{eq:cr2}),
  $N_{\rm G}$ is a number of expansion with ranges and coefficients of Gaussians.
  Hence, we have the Gaussian bases of $4N_{\rm G}$ for $F_S$ and $2N_{\rm G}$ for $F_D$ with spin-isospin dependence,
  which determine the number of variational parameters.
  For double products of $F_SF_S$, considering three-body and four-body correlations omitting two-body ladder diagram in the cluster expansion,
  number of bases becomes $2(4N_{\rm G})^2$, and $2(2N_{\rm G})^2$ for $F_DF_D$.
  In a similar way, we prepare the basis states of New-TOAMD with the Gaussian expansions of the multiple products consisting of $F_S$ and $F_D$.

  We have three categories of the correlation terms of $F$ (single), $F^2$ (double), and $F^3$ (triple) in the TOAMD wave function in Eq.~(\ref{eq:TOAMD3}), and we can use the different numbers of $N_G$ in each category, namely, $N_{G1}$ for $F$, $N_{G2}$ for $F^2$, and $N_{G3}$ for $F^3$. In the present calculation, we use the set of $(N_{G1},N_{G2},N_{G3})=(9,7,3)$ at maximum, which gives the total basis number of around 10,000 for $^4$He in Eq. (\ref{eq:linear}) considering the spin--isospin dependence of the correlation functions and the independent diagrams of the cluster expansion.
  We determine the weights of the basis states by solving the generalized eigenvalue problem given in Eq. (\ref{eq:eigen}).
  
\section{Results}\label{sec:results}

\subsection{AV6$^\prime$}

We show the results of the $s$-shell nuclei, $^3$H and $^4$He, with New-TOAMD using the AV6$^\prime$ potential not including the $LS$ force.
For $\Phi_{\rm AMD}$, we adopt the $s$-wave configurations with the centroid parameters ${\vc D}={\vc 0}$ for all nucleons.
This state is preferred in the energy minimization of each nucleus \cite{myo17a}.
Similarly, we also determine the Gaussian range parameter $\nu=0.10$ fm$^{-2}$ for $^3$H and  $\nu=0.22$ fm$^{-2}$ for $^4$He.

\begin{figure}[th]
\centering
\includegraphics[width=8.0cm,clip]{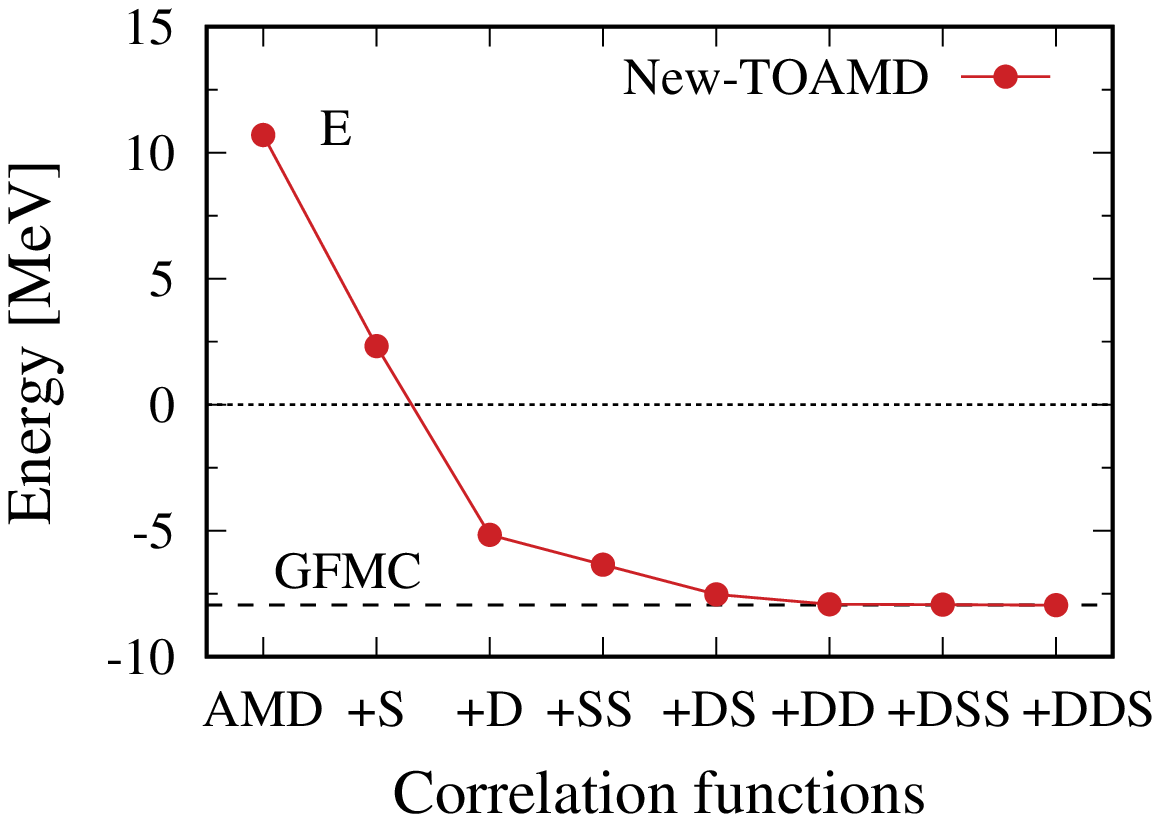}
\caption{Total energy $E$ of $^3$H with the AV6$^\prime$ potential by successively adding the correlation terms.
Dashed horizontal line is the value of GFMC.}
\label{fig:3H_AV6p_E}
\end{figure}

\begin{figure}[th]
\centering
\includegraphics[width=8.0cm,clip]{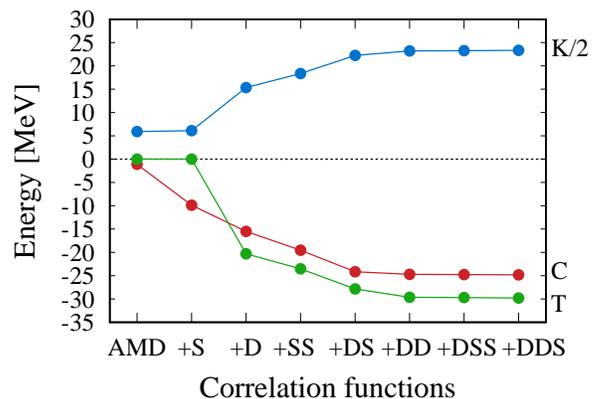}
\caption{Components of the kinetic energy (K), central force (C), and tensor force (T) in $^3$H with the AV6$^\prime$ potential.
We plot the half value of the kinetic energy as K/2.}
\label{fig:3H_AV6p_H}
\end{figure}

For $^3$H, we show the results of total energy $E$ in Table \ref{tab:3H_AV6p_E}, 
by successively adding the correlation terms in Eq. (\ref{eq:TOAMD3}).
In the notation, the labels of S and D represent $F_S$ and $F_D$, respectively,
and +S indicates the wave function of $(1+F_S)\, \Phi_{\rm AMD}$ and +DDS is the final one with $F_DF_DF_S$.
It is found that the final energy is $-7.95$ MeV, which agrees with the GFMC value of $7.95(1)$ \cite{wiringa02,wiringa_web}.
At the level of second order of TOAMD with up to $F^2$, we obtain $-7.92$ MeV, which is already close to the converging energy
and the triple correlation functions contribute by 0.03 MeV.
In Fig. \ref{fig:3H_AV6p_E}, we confirm the behaviour of the energy convergence.  
In Table \ref{tab:3H_AV6p_E}, we show the Hamiltonian components of the kinetic energy (K), central force (C), and tensor force (T).
Similar to the total energy, each Hamiltonian component almost converges at the level of second order of $F^2$,
and the triple correlation functions give the the last contributions by a few hundred keV.
In Fig. \ref{fig:3H_AV6p_H}, we show the behaviour of the convergence of the Hamiltonian components.

\begin{table}[bh]
\begin{center}
  \caption{Total energy $E$ and Hamiltonian components of the kinetic energy (K), central force (C), and tensor force (T)
    for $^3$H in New-TOAMD with the AV6$^\prime$ potential in units of MeV.
  $\nu=0.10$ fm$^{-2}$. We successively add the correlation terms.} 
\label{tab:3H_AV6p_E}
\begin{tabular}{ccccccccc}
\noalign{\hrule height 0.5pt}
        & AMD     &     +S   &  +D      &  +SS     & +DS      & +DD      &  +DSS     & +DDS \\
\noalign{\hrule height 0.5pt}
$E$& $10.70$ &  $ 2.33$  & $ -5.16$ & $ -6.35$ & $ -7.52$ &  $-7.92$ & $ -7.93$  & $-7.95$ \\ 
K  & $11.82$ &  $12.21$  & $ 30.68$ & $ 36.72$ & $ 44.50$ &  $46.43$ & $ 46.53$  & $ 46.79$ \\
C  & $-1.12$ &  $-9.89$  & $-15.52$ & $-19.53$ & $-24.16$ & $-24.71$ & $-24.77$  & $-24.88$ \\
T  & $ 0.00$ &  $ 0.00$   & $-20.32$ & $-23.53$ & $-27.86$ & $-29.64$ & $-29.70$  & $-29.86$ \\
\noalign{\hrule height 0.5pt}
\end{tabular}
\end{center}
\end{table}


We compare the previous TOAMD and the present New-TOAMD, which are
different on the treatment of the cluster expansion diagrams in the $F^2$ and $F^3$ terms in the wave function.
In the previous TOAMD, the weights of the cluster expansion diagrams of $F^2$ and $F^3$ are fixed by the symmetry factors,
while in New-TOAMD, we omit this condition of the weights and we can determine the weights of each diagram independently.
For $^3$H, there is only one diagram of [12:13] in $F^2$ and then no difference in the $F^2$ terms.
The difference comes from $F^3$ terms which have two kinds of three-body diagram of 3a and 3b as shown in Fig. \ref{fig:FFF2}.
In Table \ref{tab:3H_AV6p_CMP}, we compare the TOAMD and New-TOAMD for $^3$H, and we cannot see the difference between their energies.
This indicates that the form of $F^3$ can be kept with the symmetry factors in the wave function of $^3$H.
It is an advantage of New-TOAMD that we can evaluate the effects of the $F_DF_SF_S$ (DSS) and $F_DF_DF_S$ (DDS) individually.
It is found that both terms almost equally contribute to the total energy of $^3$H in a few hundred keV.
Among the three-body diagrams of $F^3$ shown in Fig. \ref{fig:FFF2}, contribution of 3a is important and 3b gives a minor effect.

\begin{table}[th]
\begin{center}
  \caption{Comparison of TOAMD and New-TOAMD for $^3$H with AV6$^\prime$ potential in units of MeV.
  $\nu=0.10$ fm$^{-2}$.
  We successively add the correlation diagrams to the lower rows.}
\label{tab:3H_AV6p_CMP}
\begin{tabular}{c|cccccccc}
\noalign{\hrule height 0.5pt}
        &~TOAMD    &~~New-TOAMD~\\
\noalign{\hrule height 0.5pt}
+FF     & $ -7.92$ & $- 7.92$ \\
\noalign{\hrule height 0.5pt}
+DSS,3a~& $ -$     & $- 7.93$ \\
+DSS,3b~& $ -7.93$ & $- 7.93$ \\
\noalign{\hrule height 0.5pt}
+DDS,3a~& $ -$     & $- 7.95$ \\
+DDS,3b~& $ -7.95$ & $- 7.95$ \\
\noalign{\hrule height 0.5pt}
\end{tabular}
\end{center}
\end{table}

Next, we discuss $^4$He with AV6$^\prime$ potential in the same manner as done for $^3$H.
In Table \ref{tab:4He_AV6p_E}, we show the total energy $E$ of $^4$He by successively adding the correlation terms up to DDS.
The final energy is $-25.83$ MeV, which is very close to the GFMC value of $-26.15(2)$ \cite{wiringa02,wiringa_web} by about 0.3 MeV.
This result indicates the reliability of the present New-TOAMD.
It is found that the energy contribution from the $F^3$ terms is about 0.6 MeV, which is larger than the case of $^3$H.
We also show the Hamiltonian components of $^4$He except for the Coulomb interaction, which gives the almost the constant repulsion of $0.8$ MeV
at any step of the TOAMD wave function.
The $F^3$ terms enhance each of Hamiltonian components by about a few MeV from those with up to $F^2$.

\begin{figure}[th]
\centering
\includegraphics[width=8.0cm,clip]{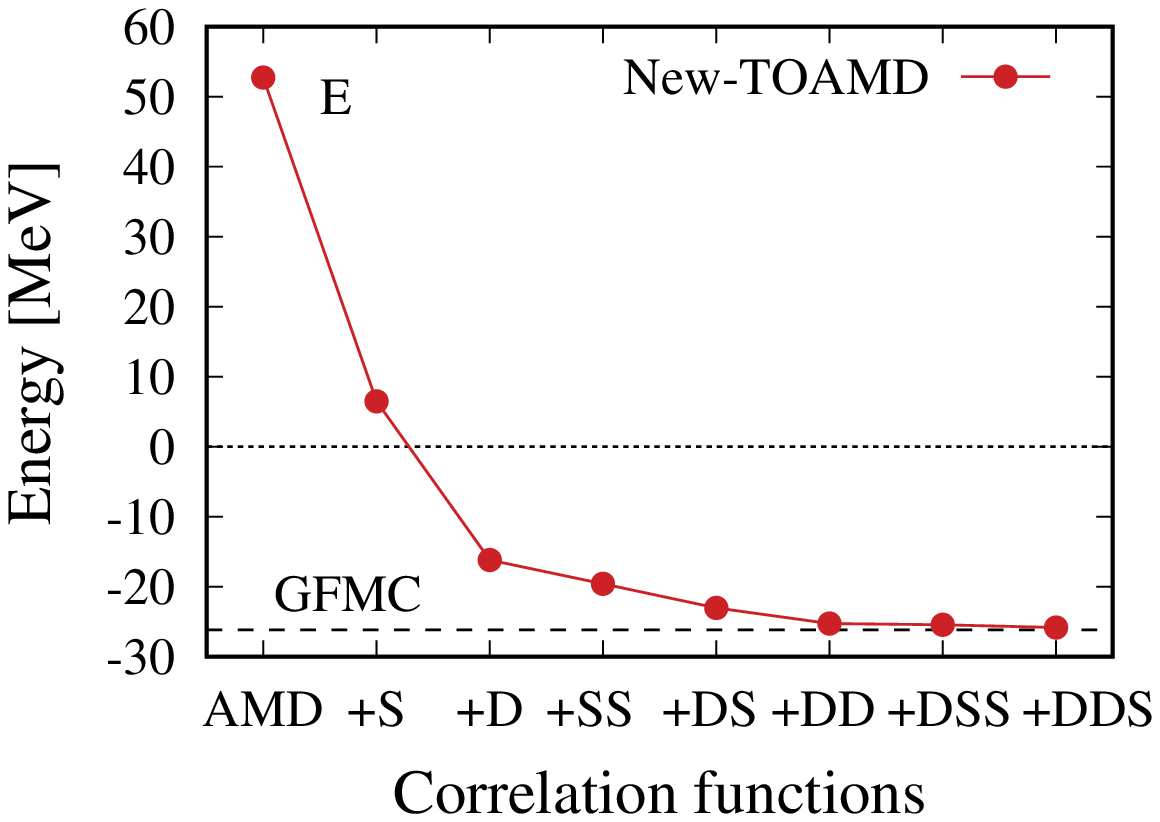}
\caption{Total energy $E$ of the ground state of $^4$He with the AV6$^\prime$ potential by successively adding the correlation terms.
Dashed horizontal line is the value of GFMC.}
\label{fig:4He_AV6p_E}
\end{figure}

\begin{figure}[th]
\centering
\includegraphics[width=8.0cm,clip]{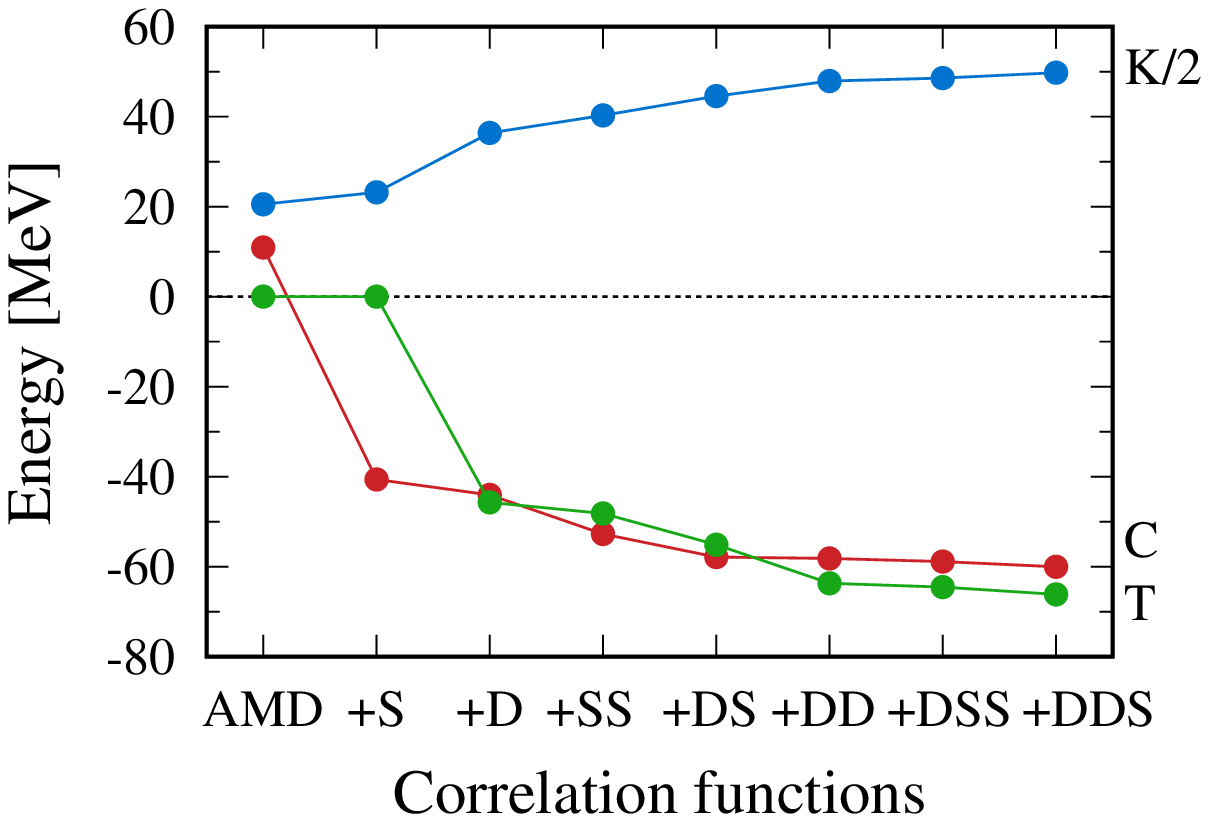}
\caption{Components of the kinetic energy (K), central force (C), and tensor force (T) in the ground state of $^4$He with the AV6$^\prime$ potential.
We plot the half value of the kinetic energy as K/2.}
\label{fig:4He_AV6p_H}
\end{figure}

In Figs. \ref{fig:4He_AV6p_E} and \ref{fig:4He_AV6p_H}, we show the convergence behaviour of the total energy $E$ and the Hamiltonian components for $^4$He.
It is found again that the total energy $E$ is approaching to the GFMC by adding the correlation terms.
We obtain the nice results with convergence at the level of second order of TOAMD and the last third order terms make the small change.
In Table \ref{tab:4He_AV6p_CMP}, we compare the results of $^4$He between the previous TOAMD and the present New-TOAMD.
In this case, we confirm the difference of 0.2 MeV in the total energy, which comes from the
independent optimization of the cluster expansion diagrams in $F^2$ and $F^3$ in New-TOAMD.
We can also see the small difference in all of the Hamiltonian components.

In Table \ref{tab:4He_AV6p_diagram}, we compare the results of $^4$He between the previous TOAMD and the present New-TOAMD
at the level of each diagram. 
The $F^2$ terms provide three-body diagram of [12:13] and four-body one [12:34], which are denoted by +FF,3 and +FF,4, respectively.
It is found at the $F^2$ level, two methods give the difference of 0.03 MeV.
Until the addition of DSS, the difference is still 0.04 MeV, but finally, it becomes 0.2 MeV by adding DDS.
This means the importance of the independent treatment of the diagrams including more tensor correlations with $F_D$.

\begin{table}[th]
\begin{center}
  \caption{Total energy $E$ and Hamiltonian components of the kinetic energy (K), central force (C), and tensor force (T)
    for $^4$He in New-TOAMD with the AV6$^\prime$ potential in units of MeV.
  $\nu=0.22$ fm$^{-2}$. We successively add the correlation terms.}
\label{tab:4He_AV6p_E}
\begin{tabular}{ccccccccc}
\noalign{\hrule height 0.5pt}
        &   AMD     &     +S    &  +D      &  +SS      & +DS       & +DD       &  +DSS   & +DDS \\
\noalign{\hrule height 0.5pt}
$E$     & $52.74$  &  $  6.49$  & $-16.19$ & $-19.58$  & $-23.02$  & $-25.27$   & $-25.44$   & $-25.83$   \\ 
K       & $41.06$  &  $ 46.41$  & $ 72.80$ & $ 80.58$  & $ 89.15$  & $ 95.83$   & $ 97.2$    & $ 99.52$ \\
C       & $10.92$  &  $-40.61$  & $-44.04$ & $-52.74$  & $-57.84$  & $-58.18$   & $-58.9$    & $-60.00$ \\
T       & $ 0.00$  &  $  0.00$  & $-45.69$ & $-48.17$  & $-55.10$  & $-63.69$   & $-64.5$    & $-66.14$ \\
\noalign{\hrule height 0.5pt}
\end{tabular}
\end{center}
\end{table}


\begin{table}[th]
\begin{center}
  \caption{Comparison of TOAMD and New-TOAMD for the Hamiltonian components of $^4$He with AV6$^\prime$ potential in units of MeV.
Radius is in units of fm.  $\nu=0.22$ fm$^{-2}$.}
\label{tab:4He_AV6p_CMP}
\begin{tabular}{c|cccccccc}
\noalign{\hrule height 0.5pt}
          &~TOAMD     &~~New-TOAMD~ \\
\noalign{\hrule height 0.5pt}
Energy    & $-25.63$  &  $-25.83$ \\
Kinetic   & $ 98.79$  &  $ 99.52$ \\
Central   & $-59.60$  &  $-60.00$ \\
Tensor    & $-65.60$  &  $-66.14$ \\
Coulomb   & $  0.79$  &  $  0.79$ \\
Radius    & $  1.46$  &  $  1.46$ \\
\noalign{\hrule height 0.5pt}
\end{tabular}
\end{center}
\end{table}

\begin{table}[th]
\begin{center}
  \caption{Comparison of TOAMD and New-TOAMD for the total energy of $^4$He with AV6$^\prime$ potential in units of MeV.
  $\nu=0.22$ fm$^{-2}$.
  We successively add the correlation diagrams to the lower rows.}
\label{tab:4He_AV6p_diagram}
\begin{tabular}{c|cccccccc}
\noalign{\hrule height 0.5pt}
        & TOAMD     &  New-TOAMD \\
\noalign{\hrule height 0.5pt}
+FF,3   & $ -$      & $-22.33$ \\
+FF,4   & $ -25.21$ & $-25.24$ \\
\noalign{\hrule height 0.5pt}
+DSS,3a & $ -$      & $-25.28$ \\
+DSS,3b & $ -$      & $-25.28$ \\
+DSS,4a & $ -$      & $-25.34$ \\
+DSS,4b & $ -$      & $-25.40$ \\
+DSS,4c & $ -$      & $-25.44$ \\
+DSS,4d & $-25.40$  & $-25.44$ \\
\noalign{\hrule height 0.5pt}
+DDS,3a & $ -$      & $-24.48$ \\
+DDS,3b & $ -$      & $-25.48$ \\
+DDS,4a & $ -$      & $-25.52$ \\
+DDS,4b & $ -$      & $-25.65$ \\
+DDS,4c & $ -$      & $-25.79$ \\
+DDS,4d & $-25.63$  & $-25.83$ \\
\noalign{\hrule height 0.5pt}
\end{tabular}
\end{center}
\end{table}

\subsection{AV8$^\prime$}

We show the results of the $s$-shell nuclei, $^3$H and $^4$He using the AV8$^\prime$ potential including the $LS$ force.
We use the centroid parameters ${\vc D}={\vc 0}$ in the AMD wave function $\Phi_{\rm AMD}$ for all nucleons
and the Gaussian range parameters are $\nu=0.10$ fm$^{-2}$ for $^3$H and  $\nu=0.22$ fm$^{-2}$ for $^4$He.
They are the same conditions as obtained in the case of the AV6$^\prime$ potential.

\begin{figure}[th]
\centering
\includegraphics[width=8.0cm,clip]{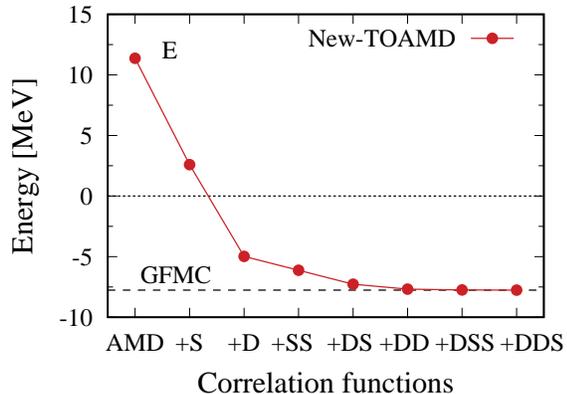}
\caption{Total energy $E$ of the ground state of $^3$H with the AV8$^\prime$ potential by successively adding the correlation terms.
Dashed horizontal line is the value of GFMC.}
  \label{fig:3H_AV8p_E}
\end{figure}

\begin{figure}[th]
\centering
\includegraphics[width=8.0cm,clip]{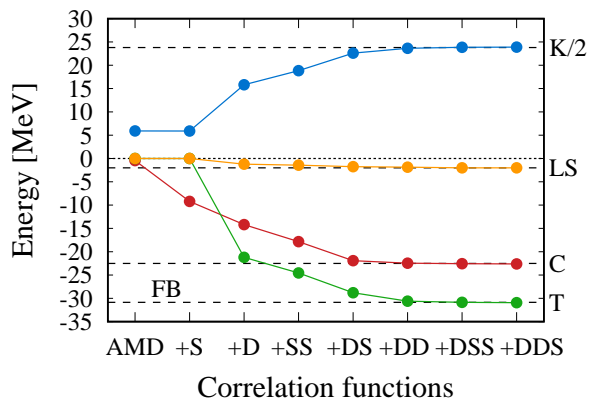}
\caption{Components of the kinetic energy (K), central force (C), tensor force (T), $LS$ force in the ground state of $^3$H with the AV8$^\prime$ potential.
  We plot the half value of the kinetic energy as K/2. 
  Dashed horizontal lines are the values of the few-body calculation (FB) taken from Ref. \cite{suzuki08}.
}
\label{fig:3H_AV8p_H}
\end{figure}

We show the results of total energy $E$ of $^3$H in Table \ref{tab:3H_AV8p_E}. 
The final value is $-7.76$ MeV, which agrees with the GFMC value of $7.76(1)$ \cite{wiringa02,wiringa_web}.
At the second order of TOAMD within $F^2$, we obtain $-7.68$ MeV, which gives the difference of 0.08 MeV
and already very close to the converging energy.
The $F^3$ terms contribute to 0.08 MeV in the energy,
and this value is similar to the results using the AV6$^\prime$ potential as shown in Table \ref{tab:3H_AV6p_E}. 
In Fig. \ref{fig:3H_AV8p_E}, we confirm the behaviour of the energy convergence.  
In Table \ref{tab:3H_AV8p_E}, we show the Hamiltonian components in the same notation used for the AV6$^\prime$ potential
and the $LS$ component is added.
Similar to the total energy, each Hamiltonian component almost converges at the level of second order of $F^2$.
It is found that the $F^3$ terms work to enhance every component.
In Fig. \ref{fig:3H_AV8p_H}, we show the behaviour of the convergence of the Hamiltonian components.
From the table, in the effect of the $F^3$ terms, the tensor force changes by 0.3 MeV, which is larger than that of the central force by 0.1 MeV.
This indicates the importance of tensor correlation in the third order of TOAMD.

\begin{table}[th]
\begin{center}
  \caption{Total energy $E$ and Hamiltonian components of the kinetic energy (K), central force (C), tensor force (T), and LS force (LS)
   for $^3$H in New-TOAMD with the AV8$^\prime$ potential in units of MeV.
  $\nu=0.10$ fm$^{-2}$. We successively add the correlation terms.}
\label{tab:3H_AV8p_E}
\begin{tabular}{ccccccccc}
\noalign{\hrule height 0.5pt}
     &  AMD     & +S       &  +D       &  +SS     &  +DS     & +DD     & +DSS     & +DDS    \\
\noalign{\hrule height 0.5pt}
$E$  & $11.37$  & $2.60$   & $ -4.98$  & $ -6.12$  & $ -7.27$  &  $-7.68$  & $-7.75$   & $-7.76$ \\ 
K    & $11.82$  &  $11.80$ & $ 31.64$  & $ 37.69$  & $ 45.22$  & $ 47.27$  & $47.70$   & $ 47.80$ \\
C    & $-0.45$  &  $-9.20$ & $-14.19$  & $-17.83$  & $-21.93$  & $-22.47$  & $-22.58$  & $-22.62$ \\
T    & $ 0.00$  &  $0.00$  & $-21.22$  & $-24.56$  & $-28.81$  & $-30.61$  & $-30.87$  & $-30.93$ \\
LS   & $ 0.00$  &  $0.00$  & $ -1.21$  & $ -1.34$  & $ -1.75$  & $ -1.86$  & $ -2.00$  & $ -2.00$ \\
\noalign{\hrule height 0.5pt}
\end{tabular}
\end{center}
\end{table}

We compare the previous TOAMD and the present New-TOAMD.
In Table \ref{tab:3H_AV8p_CMP}, we compare the TOAMD and New-TOAMD for $^3$H, and we cannot see the difference between their energies.
This is the same conclusion as obtained with the AV6$^\prime$ potential.
We evaluate the effects of DSS and DDS successively, and both terms contribute to the total energy of $^3$H.
It is noted that when we consider the DDS term only in $F^3$, the energy is obtained as $-7.75$ MeV, which is the same value
of the DSS case. Hence these two terms are considered to equally contribute to the $^3$H solutions.

\begin{table}[th]
\begin{center}
  \caption{Comparison of TOAMD and New-TOAMD for $^3$H with AV8$^\prime$ potential in units of MeV.
  $\nu=0.10$ fm$^{-2}$.
  We successively add the correlation diagrams to the lower rows.}
\label{tab:3H_AV8p_CMP}
\begin{tabular}{c|cccccccc}
\noalign{\hrule height 0.5pt}
        &~TOAMD      &~~New-TOAMD~ \\
\noalign{\hrule height 0.5pt}
+FF     & $ -7.68$   & $- 7.68$ \\
\noalign{\hrule height 0.5pt}
+DSS,3a & $ -$       & $- 7.75$ \\
+DSS,3b & $ -7.75$   & $- 7.75$ \\
\noalign{\hrule height 0.5pt}
+DDS,3a & $ -$       & $- 7.76$ \\
+DDS,3b & $ -7.76$   & $- 7.76$ \\
\noalign{\hrule height 0.5pt}
\end{tabular}
\end{center}
\end{table}

Next, we discuss $^4$He with the AV8$^\prime$ potential.
In Table \ref{tab:4He_AV8p_E}, we show the total energy $E$ of $^4$He by successively adding the correlation terms up to DDS.
The final value is $-24.83$ MeV, which is close to the GFMC value of $-25.14(2)$ by about 0.3 MeV \cite{wiringa02,wiringa_web}.
This accuracy is the same as obtained in the AV6$^\prime$ potential and we confirm the reliability of the present New-TOAMD.
In the table, we also show the Hamiltonian components of $^4$He except for the Coulomb interaction, which gives the repulsion energy of $0.8$ MeV.

\begin{figure}[th]
\centering
\includegraphics[width=8.0cm,clip]{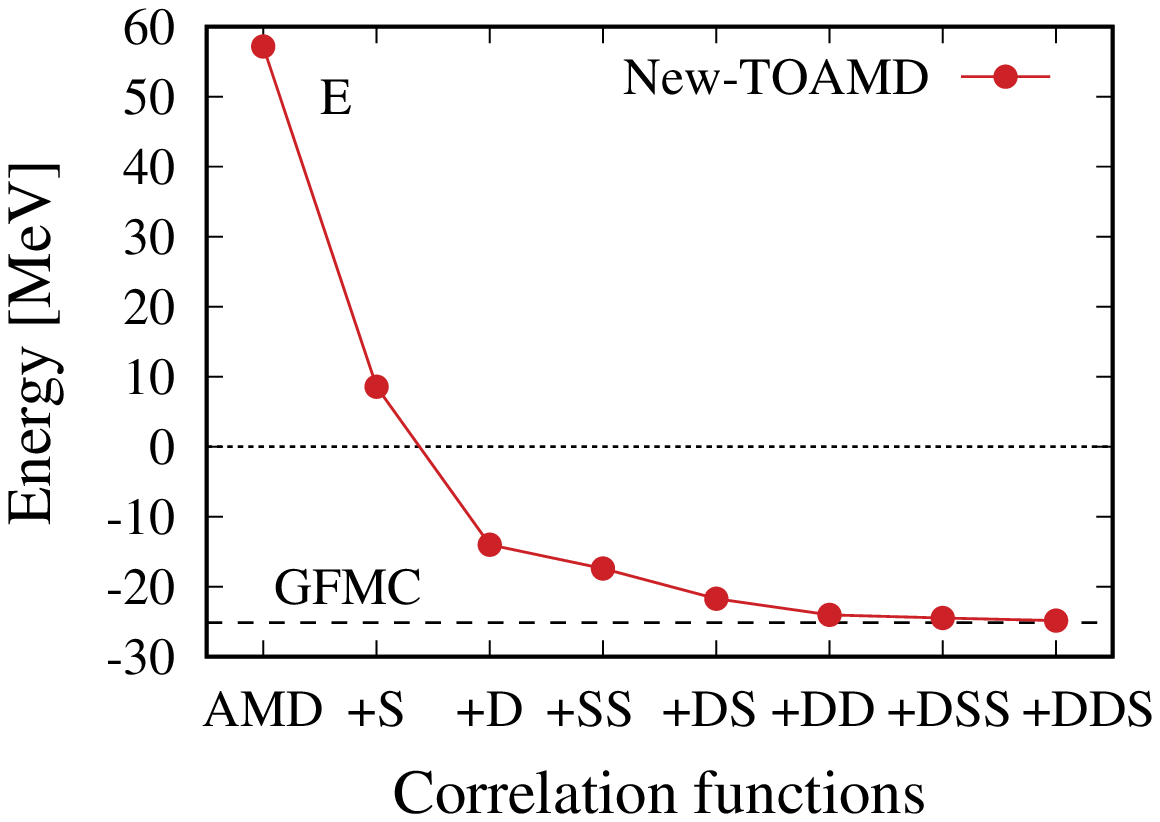}
\caption{Total energy $E$ of the ground state of $^4$He with the AV8$^\prime$ potential by successively adding the correlation terms.
Dashed horizontal line is the value of GFMC.}
  \label{fig:4He_AV8p_E}
\end{figure}

\begin{figure}[th]
\centering
\includegraphics[width=8.0cm,clip]{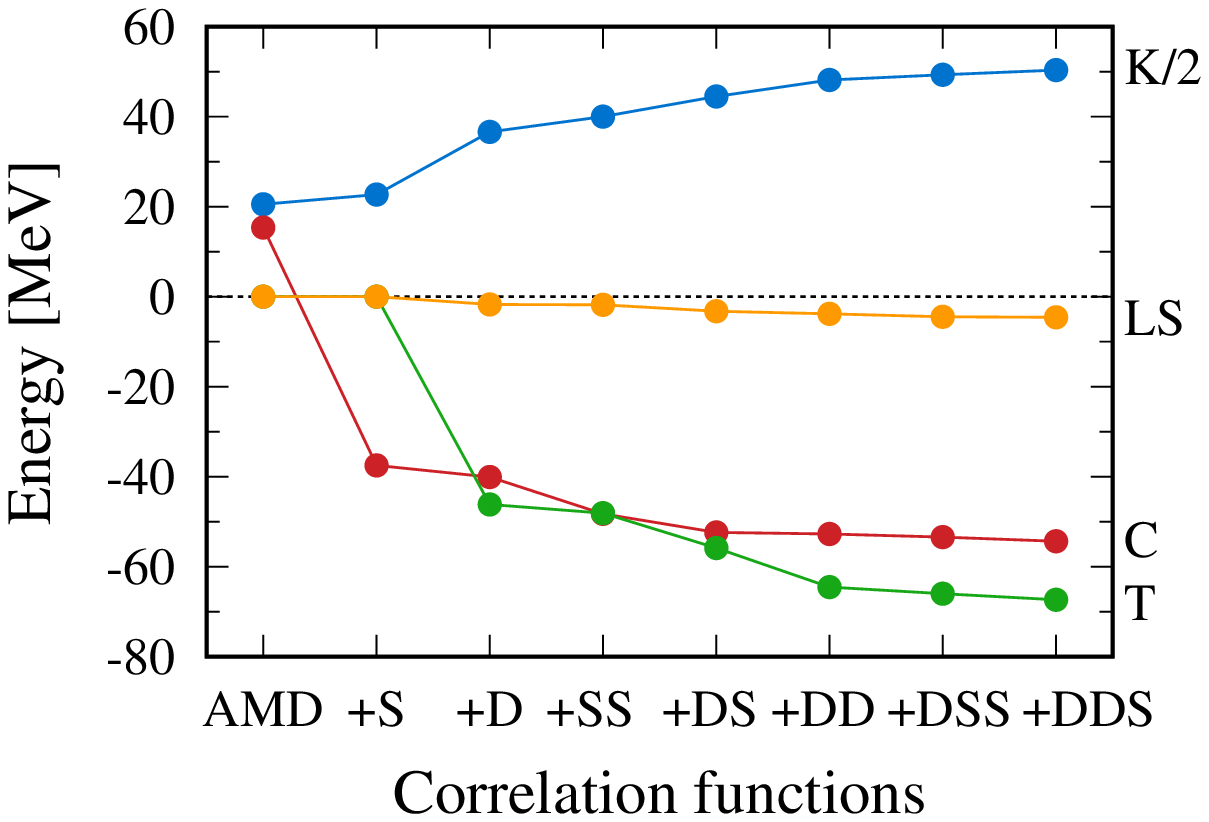}
\caption{Components of the kinetic energy (K), central force (C), tensor force (T), $LS$ force in the ground state of $^4$He with the AV8$^\prime$ potential.
We plot the half value of the kinetic energy as K/2.}
\label{fig:4He_AV8p_H}
\end{figure}

In Figs. \ref{fig:4He_AV8p_E} and \ref{fig:4He_AV8p_H},
we show the convergence behaviour of the total energy $E$ and the Hamiltonian components.
It is found again that the total energy $E$ is approaching to the GFMC by adding the correlation terms.
We obtain the nice results with convergence at the level of second order of TOAMD and the last third order terms make the small change.
For the effect of the third order by adding the $F^3$ terms,
in the interaction energy, the tensor force changes by 2.9 MeV, which is larger than that of the central force by 1.6 MeV.
This trend is similar to the results of $^3$H, and indicates the importance of tensor correlation.

In Table \ref{tab:4He_AV8p_CMP}, we compare the results of $^4$He between the previous TOAMD and the present New-TOAMD.
The energy difference is obtained as 0.23 MeV ,which is a similar value to the AV6$^\prime$ case. 
This difference is due to the independent treatment of the diagrams in $F^2$ and $F^3$ in New-TOAMD.
We can also see the difference in the Hamiltonian components.
For central and tensor forces, both components increase in New-TOAMD.

In Table \ref{tab:4He_AV8p_diagram}, we compare the total energies of $^4$He between the previous TOAMD and the present New-TOAMD
at each diagram, which is successively added.
It is found at the $F^2$ level, two methods give the difference of 0.01 MeV.
By adding DSS, the energy difference becomes large as 0.08 MeV, and finally, it becomes 0.23 MeV.
This indicates the importance of DDS term having more tensor correlations by $F_D$.
This is the same tendency confirmed in the case of the AV6$^\prime$ potential.

\begin{table}[th]
\begin{center}
  \caption{Total energy $E$ and Hamiltonian components of the kinetic energy (K), central force (C), tensor force (T), and LS force (LS)
   for $^4$He in New-TOAMD with the AV8$^\prime$ potential in units of MeV.
  $\nu=0.22$ fm$^{-2}$. We successively add the correlation terms.}
\label{tab:4He_AV8p_E}
\begin{tabular}{ccccccccc}
\noalign{\hrule height 0.5pt}
       & AMD    &     +S    &  +D       &  +SS    & +DS     & +DD     &  +DSS   & +DDS \\
\noalign{\hrule height 0.5pt}
$E$  & $57.18$  &  $  8.56$ & $-14.02$  &   $-17.40$  & $-21.72$  & $-24.02$   & $-24.46$    & $-24.83$  \\ 
K    & $41.06$  &  $ 45.35$ & $ 73.22$  &   $ 80.01$  & $ 88.97$  & $ 96.29$   & $ 98.65$    & $100.70$ \\
C    & $15.37$  &  $-37.48$ & $-40.07$  &   $-48.25$  & $-52.36$  & $-52.74$   & $-53.42$    & $-54.35$ \\
T    & $ 0.0$   &  $  0.0$  & $-46.16$  &   $-48.11$  & $-55.85$  & $-64.51$   & $-65.99$    & $-67.36$ \\
LS   & $ 0.0$   &  $  0.0$  & $ -1.74$  &   $- 1.80$  & $ -3.23$  & $- 3.82$   & $ -4.47$    & $- 4.58$ \\
\noalign{\hrule height 0.5pt}
\end{tabular}
\end{center}
\end{table}

\begin{table}[th]
\begin{center}
  \caption{Comparison of TOAMD and New-TOAMD for $^4$He with AV8$^\prime$ potential in units of MeV.
  $\nu=0.22$ fm$^{-2}$.}
\label{tab:4He_AV8p_CMP}
\begin{tabular}{c|cccccccc}
\noalign{\hrule height 0.5pt}
         &~TOAMD     &~~New-TOAMD~ \\
\noalign{\hrule height 0.5pt}
Energy   & $-24.60$  & $-24.83$ \\
Kinetic  & $ 99.54$  & $100.70$ \\
Central  & $-53.89$  & $-54.35$ \\
Tensor   & $-66.62$  & $-67.36$ \\
LS       & $ -4.40$  & $ -4.58$ \\
Coulomb  & $  0.78$  & $  0.77$ \\
\noalign{\hrule height 0.5pt}
\end{tabular}
\end{center}
\end{table}

\begin{table}[th]
\begin{center}
  \caption{Comparison of TOAMD and New-TOAMD for $^4$He with AV8$^\prime$ potential in units of MeV.
  $\nu=0.22$ fm$^{-2}$.
  We successively add the correlation diagrams to the lower rows.}
\label{tab:4He_AV8p_diagram}
\begin{tabular}{c|cccccccc}
\noalign{\hrule height 0.5pt}
      & TOAMD     &  New-TOAMD \\
\noalign{\hrule height 0.5pt}
+FF,3 & $ - $       & $-21.14$ \\
+FF,4 & $ -24.01$   & $-24.02$ \\
\noalign{\hrule height 0.5pt}
+DSS,3a & $ -$     & $-24.22$ \\
+DSS,3b & $ -$     & $-24.22$ \\
+DSS,4a & $ -$     & $-24.29$ \\
+DSS,4b & $ -$     & $-24.39$ \\
+DSS,4c & $ -$     & $-24.46$ \\
+DSS,4d & $-24.38$ & $-24.46$ \\
\noalign{\hrule height 0.5pt}
+DDS,3a & $ -$     & $-24.49$ \\
+DDS,3b & $ -$     & $-24.49$ \\
+DDS,4a & $ -$     & $-24.54$ \\
+DDS,4b & $ -$     & $-24.65$ \\
+DDS,4c & $ -$     & $-24.79$ \\
+DDS,4d & $-24.60$ & $-24.83$ \\
\noalign{\hrule height 0.5pt}
\end{tabular}
\end{center}
\end{table}

\begin{figure}[tbh]
\centering
\includegraphics[width=8.0cm,clip]{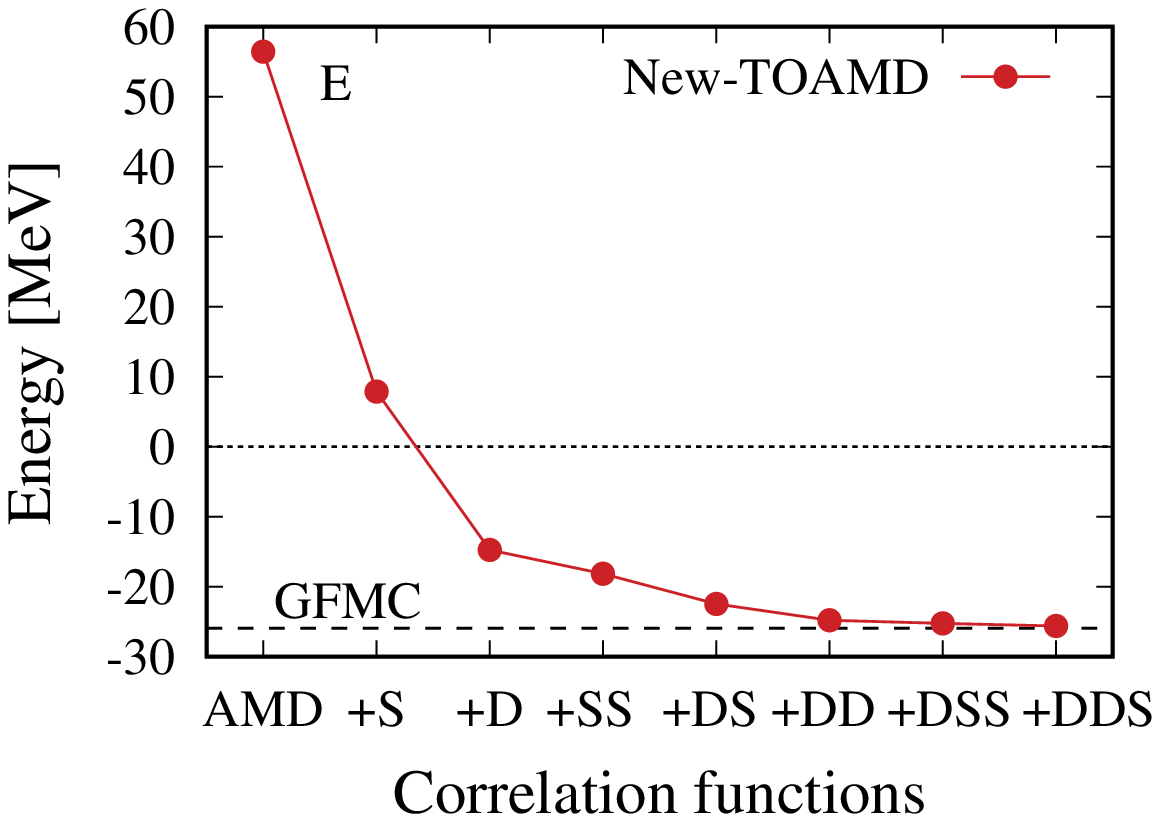}
\caption{Total energy $E$ of the ground state of $^4$He with the AV8$^\prime$ potential without the Coulomb interaction by successively adding the correlation terms.
Dashed horizontal line is the value of GFMC.}
\label{fig:4He_AV8p_E_noClm}
\end{figure}

\begin{figure}[tbh]
\centering
\includegraphics[width=8.0cm,clip]{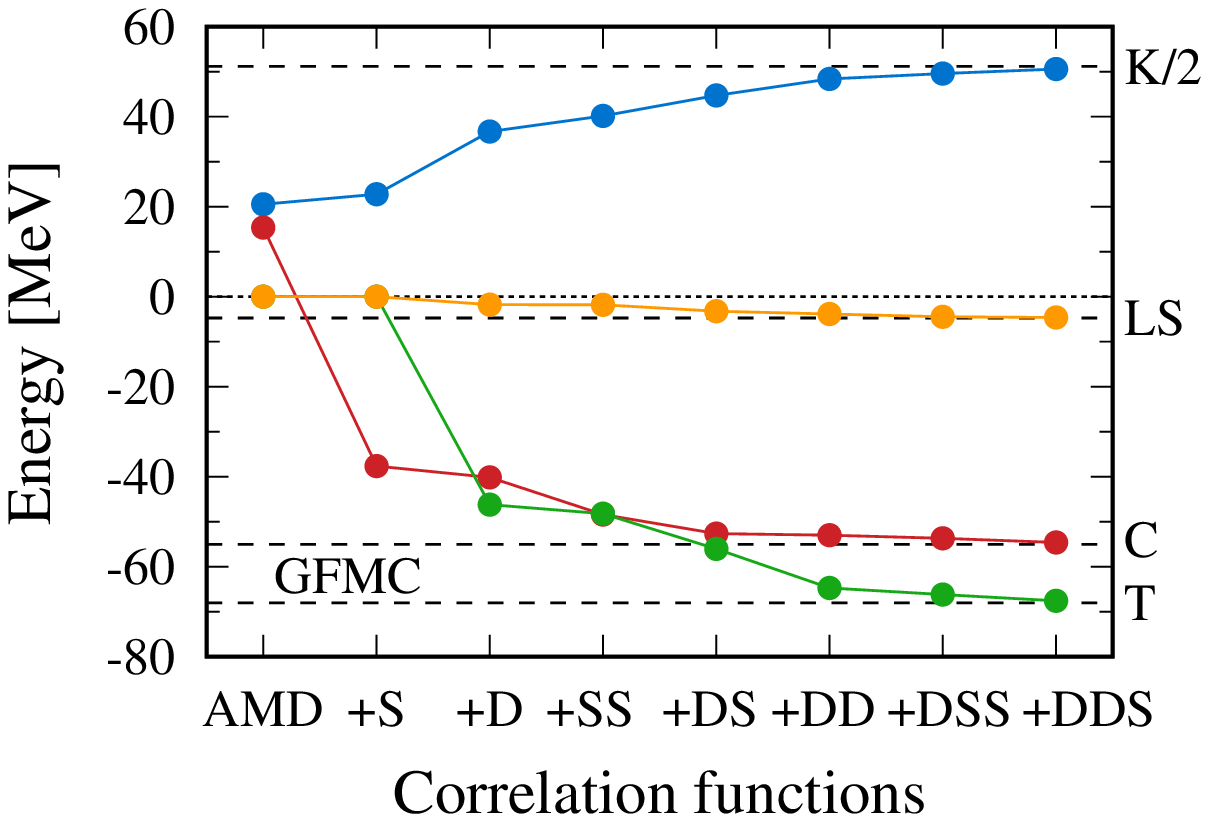}
\caption{Components of the kinetic energy (K), central force (C), tensor force (T) , $LS$ force in the ground state of $^4$He with the AV8$^\prime$ potential without the Coulomb interaction.
We plot the half value of the kinetic energy as K/2.
Dashed horizontal lines are the values of GFMC \cite{kamada01}.}
\label{fig:4He_AV8p_H_noClm}
\end{figure}

Finally, we compare the results of $^4$He in New-TOAMD with those of benchmark calculations 
using AV8$^\prime$ potential without the Coulomb interaction \cite{kamada01}.
We select two calculations of GFMC and the Faddeev-Yakubovsky (FY).
The results are shown in Table \ref{tab:4He_AV8p_benchmark}
and in New-TOAMD, difference of the total energy is about 0.3 MeV, which is the same value obtained
with the Coulomb interaction as shown in Table \ref{tab:4He_AV8p_E}.
For each Hamiltonian component we can confirm the good agreement between those of New-TOAMD and other calculations
within about 1 MeV. Radius also shows a very similar value.
In Figs. \ref{fig:4He_AV8p_E_noClm} and  \ref{fig:4He_AV8p_H_noClm}, we show the convergence of total energy
and the Hamiltonian components in New-TOAMD by successively adding the correlation terms until the DDS term.
We can clear confirm that the New-TOAMD solutions approaches to the values of GFMC with good convergence.

From these comparisons, it is concluded that variational accuracy increases from TOAMD to New-TOAMD to describe the $NN$ correlations.
These results indicate that the independent treatment of the cluster expansion diagrams works successfully.
It is an interesting subject to apply the present New-TOAMD to the calculation of the $p$-shell nuclei.

\begin{table}[th]
\begin{center}
  \caption{Comparison of TOAMD and New-TOAMD for $^4$He with AV8$^\prime$ potential without Coulomb interaction.
    Energies are in units of MeV. Radius is in units of fm. $\nu=0.22$ fm$^{-2}$.
    The values of GFMC and FY are taken from Ref. \cite{kamada01}.}
  \label{tab:4He_AV8p_benchmark}
\begin{tabular}{c|cc|cccccc}
\noalign{\hrule height 0.5pt}
        &~TOAMD     &~New-TOAMD~ &  GFMC    & FY \\
\noalign{\hrule height 0.5pt}
Energy  & $-25.37$  & $-25.61$   & $-25.93(2)$  & $-25.94(5)$ \\
Kinetic & $100.04$  & $101.23$   & $102.3(1.0)$ & $102.39(5)$ \\
Central & $-54.15$  & $-54.63$   & $-55.05(70)$ & $-55.26$ \\
Tensor  & $-66.85$  & $-67.61$   & $-68.05(70)$ & $-68.35$ \\
 LS     & $ -4.42$  & $ -4.60$   & $ -4.75(5)$  & $ -4.72$ \\
Radius  & $ 1.485$  & $ 1.478$   & $  1.490(5)$ & $1.485(3)$
\\
\noalign{\hrule height 0.5pt}
\end{tabular}
\end{center}
\end{table}

\section{Summary}\label{sec:summary}
We developed a new variational method of ``tensor-optimized antisymmetrized molecular dynamics'' (TOAMD) for finite nuclei,
which is a successive approach to treat bare nucleon-nucleon interactions directly.
In TOAMD, we introduce the two-body correlation functions with tensor and central-types and further consider the multiple products consisting of them.
The multiple products of the correlation functions are expressed into the series of the many-body diagrams using the cluster expansion.
The forms of the correlation functions are determined in the total-energy variation.

In this paper, we extend TOAMD in two aspects:
1) We include the triple correlation functions $F^3$ as the third order of TOAMD.
2) We treat independently each of the cluster expansion diagrams and determine the correlation functions in each diagram variationally. We call this new method New-TOAMD.
From the second aspect, we can also omit the duplicated diagrams, the physical effect of which can be represented by other diagrams.
These improvements extend the variational space of TOAMD and increase the variational accuracy using bare nucleon-nucleon interaction.
We calculated the total energy and Hamiltonian components of $s$-shell nuclei, and
confirmed that the present improvements make the solutions of TOAMD get close or identical to those of the few-body calculations.
It is interesting in future to apply the present New-TOAMD to the $p$-shell and larger mass-number nuclei,
in which we can determine all the correlation functions based on the variational principle.

Three-nucleon force is also an interesting subject to be investigated in nuclear structure.
In the three-nucleon force, multi-pion exchange process is an important component, which results in the multiples of the central and tensor operators.
This property has an analogy with the cluster expansion of the products of the correlation functions in TOAMD, as is shown in the three-body diagram of [12:13] in Fig. (\ref{fig:FF2}).
Hence, the framework of TOAMD is straightforward to treat the three-nucleon force and some useful formulae are given in Ref.~\cite{myo15}.
It is interesting to investigate the three-body force effect in nuclei in TOAMD.

\section*{Acknowledgments}
This work was supported by JSPS KAKENHI Grants No. JP18K03660.
Numerical calculations were partly achieved through the use of OCTOPUS at the Cybermedia Center, Osaka University.

\section*{References}
\def\JL#1#2#3#4{ {{\rm #1}} \textbf{#2}, #3 (#4).}  
\nc{\PR}[3]     {\JL{Phys. Rev.}{#1}{#2}{#3}}
\nc{\PRC}[3]    {\JL{Phys. Rev.~C}{#1}{#2}{#3}}
\nc{\PRA}[3]    {\JL{Phys. Rev.~A}{#1}{#2}{#3}}
\nc{\PRL}[3]    {\JL{Phys. Rev. Lett.}{#1}{#2}{#3}}
\nc{\NP}[3]     {\JL{Nucl. Phys.}{#1}{#2}{#3}}
\nc{\NPA}[3]    {\JL{Nucl. Phys.}{A#1}{#2}{#3}}
\nc{\PL}[3]     {\JL{Phys. Lett.}{#1}{#2}{#3}}
\nc{\PLB}[3]    {\JL{Phys. Lett.~B}{#1}{#2}{#3}}
\nc{\PTP}[3]    {\JL{Prog. Theor. Phys.}{#1}{#2}{#3}}
\nc{\PTPS}[3]   {\JL{Prog. Theor. Phys. Suppl.}{#1}{#2}{#3}}
\nc{\PTEP}[3]   {\JL{Prog. Theor. Exp. Phys.}{#1}{#2}{#3}}
\nc{\PRep}[3]   {\JL{Phys. Rep.}{#1}{#2}{#3}}
\nc{\PPNP}[3]   {\JL{Prog.\ Part.\ Nucl.\ Phys.}{#1}{#2}{#3}}
\nc{\JPG}[3]    {\JL{J. of Phys. G}{#1}{#2}{#3}}
\nc{\CPC}[3]    {\JL{Chin. Phys. C}{#1}{#2}{#3}}
\nc{\andvol}[3] {{\it ibid.}\JL{}{#1}{#2}{#3}}

\end{document}